\documentclass[conference]{IEEEtran}
\IEEEoverridecommandlockouts

\usepackage{cite}
\usepackage{amsmath,amssymb,amsfonts}
\usepackage{algorithmic}
\usepackage{algorithm}
\usepackage{graphicx}
\usepackage{textcomp}
\usepackage{xcolor}
\usepackage[left=.625in,right=.625in,top=.751in,bottom=1.05in]{geometry}
\def\BibTeX{{\rm B\kern-.05em{\sc i\kern-.025em b}\kern-.08em
    T\kern-.1667em\lower.7ex\hbox{E}\kern-.125emX}}
\usepackage[hidelinks]{hyperref}

\begin{document}

\title{Sustainable Grid through Distributed Data Centers \\
\vspace{1mm}
\large{Spinning AI Demand for Grid Stabilization and Optimization}\thanks{© 2025 IEEE. Personal use of this material is permitted. Permission from IEEE must be obtained for all other uses, in any current or future media, including reprinting/republishing this material for advertising or promotional purposes, creating new collective works, for resale or redistribution to servers or lists, or reuse of any copyrighted component of this work in other works.}}


\author{
  \IEEEauthorblockN{
    Scott C Evans$^1$,
    Nathan Dahlin$^3$,
    Ibrahima Ndiaye$^2$,
    Sachini Piyoni Ekanayake$^1$,\\
    Alexander Duncan$^1$,
    Blake Rose$^1$,
    Hao Huang$^1$
  }
  \IEEEauthorblockA{
    \textit{GE Vernova Advanced Research Center: } $^1$\textit{AI-Machine Learning, Robotics Lab; }$^2$\textit{Electrification Mission} \\
          $^3$\textit{ECE Department, University at Albany SUNY} \\
    \textit{$^1$evans@ge.com,
    $^3$ndahlin@albany.edu,$^2$Ndiaye@ge.com, $^1$Sachini.Ekanayake@ge.com,} \\ \textit{$^1$Alexander.K.Duncan@ge.com, $^1$Blake.Rose@ge.com, $^1$Hao.Huang@ge.com}
  }
}

\maketitle

\begin{abstract}
We propose a disruptive paradigm to actively place and schedule TWhrs of parallel AI jobs strategically on the grid, at distributed, grid-aware high performance compute data centers (HPC) capable of using their massive power and energy load to stabilize the grid while reducing grid build-out requirements, maximizing use of renewable energy, and reducing Green House Gas (GHG) emissions.  Our approach will enable the creation of new, value adding markets for spinning compute demand, providing market based incentives that will drive the joint optimization of energy and learning. \end{abstract}


\section{Introduction}
Data center HPC energy demand totaled 200TWh (4\% of US electricity) in 2022 and is expected to grow to 260TWh (6\%) by 2026 \cite{b1} and to 9.1\% of US electricity consumption by 2030 \cite{b2}. This growth trends in concentrated areas, creating uneven geographic distribution and forcing significant expansion where green generation resources may not be abundant, affordable, or optimally used. Buildout to energize this +2x demand growth under current paradigms is projected to require US grid expansion by +1x at insurmountable costs (at least +4x) \cite{b3}. In this paper, we describe a paradigm of distributed HPC that leverages the near zero cost of moving information over the internet, enabling the seamless routing of energy intensive, massively parallelizable HPC Monte-Carlo (HPCMC) and other high value AI jobs to available green energy capacity. This process will stabilize the grid, cut grid build-out requirements by half, and maximize the use of renewable, GHG-free energy.  

\begin{figure}[!htbp]
\centering
\includegraphics[width=\linewidth]{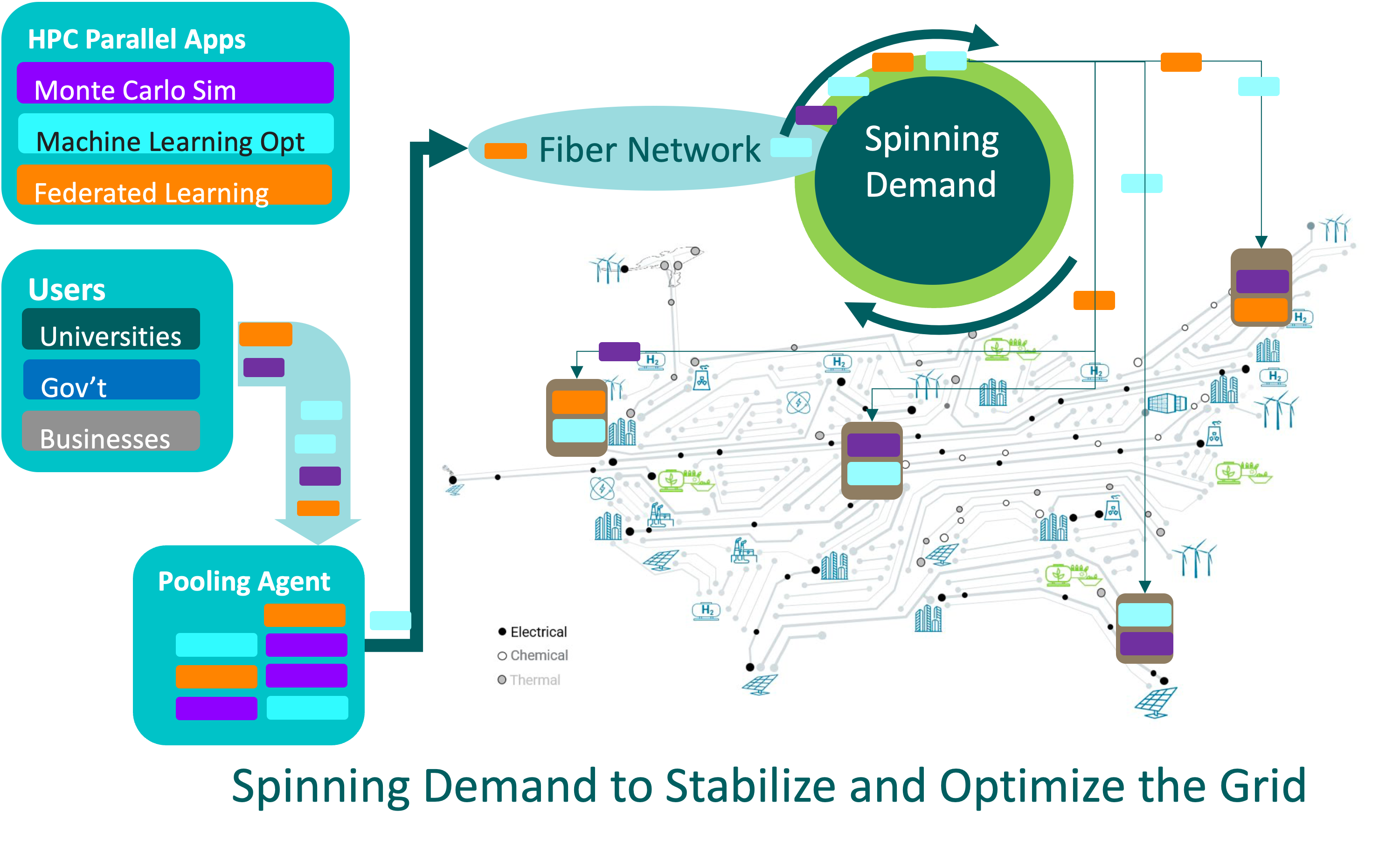}
\caption{Vision for Sustainable Grid through distributed data centers:  spinning demand of batch-able compute jobs distributed to stabilize the grid.  }
\label{figSpinning}
\end{figure}

Our approach provides grid-centric, disruptive methods to place and schedule critical portions of HPC loads, allowing lower peak demand requirements for grid capacity and reliability while minimizing GHG emissions.  
Our vision of a sustainable grid through ``spinning demand" served to distributed data centers is shown in Fig. \ref{figSpinning}, where a pooling agent distributes a backlog of HPCMC jobs throughout the grid to level load, maximize renewables, and stabilize the grid.  

Our paper is structured as follows: After reviewing previous work, we describe our approach to providing grid-centric disruptive methods to prescribe and schedule a critical portion of HPC loads.
We then present simulation results that illustrate the efficacy of our paradigm, followed by analysis of market operation and pricing, and cost-benefit analysis.

The key contributions of this work include: (i) A novel grid-centric paradigm focused on the distribution of batchable compute loads. (ii) An algorithm and accompanying simulations demonstrating benefits including reductions in renewable curtailment and required spinning reserve, with particular attention to the impact of HPC and renewable co-location. (iii) Assessment of the sensitivity of economic benefits of our approach to factors including end user flexibility, degree of co-location of HPC and renewable resources, and renewable supplier market participation rules. 

\section{Prior Work}
Our vision builds upon the state of art, such as “Zero Carbon Grid” \cite{b7} paradigms and related technologies, by jointly optimizing the electric and HPC loads, making possible solutions previously not considered. Today, grid stabilization is achieved via costly high-capacity capacitors, synchronous power condensers, and the overprovisioning of spinning reserve and the grid capacity.  We leverage theoretical concepts put forward in GE Vernova’s “Green AI” and “Entropy Economy” initiatives \cite{b8} that show promise for the joint optimization of computational learning and energy, and harvest these concepts with a grid centric view.  Specifically, \cite{optimizing_2023} describes how to optimally schedule energy intensive ML tasks in a distributed way by allocating load to where the free energy exists, such that the priority of the task completion is also achieved. 

In recent years the adoption of deep learning models has grown rapidly due to their superior performance in areas like natural language processing (e.g., large language models (LLMs)). At the same time, the complexity of these models has grown at a comparable pace, such that both training and inference require higher energy consumptions \cite{minaee2024large, chang2024survey, yarally2023uncovering}. 
In this setting, \cite{huang2024energy} proposed a streaming neural network that uses incremental weighted Principal Component Analysis (PCA),
facilitating swift and precise classification tasks given restricted memory and energy resources while ensuring high processing speed.
Further, to amplify learning while jointly minimizing energy and carbon costs, \cite{evans2024entropy} introduces a concept of ``Additive AI", designed to harmonize with the Kolmogorov Complexity and Kolmogorov Structure Function. 
 
At the same time, tracking CO2 emission from various machine learning models has emerged \cite{budennyy2022eco2aicarbonemissionstracking}. For instance,  \cite{ukarande2024pact} presents an accurate system-wide power analysis and carbon emission tracking method called ``PACT'' which measures the total power consumed by hardware resources when running benchmark machine learning tasks.
To this end, \cite{konopik2023fundamental} discusses the advantage of finite-time parallel computing over serial computing using insights from non-equilibrium thermodynamics. In parallel to the advancement of LLMs, \cite{McDonald2022GreatPG, argerich2024measuring} present techniques to measure and reduce energy consumption for training and inference for language models including avenues like reducing carbon impact. Further, \cite{alizadeh2024green} explores the impact of energy efficiency of inference in deep learning runtime infrastructure by comparing different execution providers using a single model i.e., ResNet.  Since many LLMs are trained once and used many times later, \cite{desislavov2021compute}
analyses the inference costs in the applications of computer vision and natural language processing in the aspect of  AI and energy consumption.

On the other hand, it is important to focus on how this energy increment affects the power grid. To this end, \cite{li2024unseen} emphasizes the impact on grid reliability due to the addition of load from tied to the exponential growth in machine learning models, while \cite{Sourceability} emphasizes the impact of AI on power demand and need for power distribution and management (e.g., energy storage systems and smart grid technologies) to support the growing AI industry.  Regarding state-of-the art methods deployed today, \cite{zhang2025optimal} describes development of battery energy storage systems to support variability introduce in the grid due to renewables and \cite{qays2025h} describes the synchronous condenser usage for weak grids. Still, these existing grid stability methods do not consider explicitly leveraging the emerging data center energy demand \cite{b1}. 

\section{Proposed Paradigm}
\label{sec:Proposed Paradigm}
\subsection{Parallelization of Computational jobs}\label{sec:parallel_HPCMC}
Traditionally, HPC vendors have required uninterruptible power and energy supply to support their ever-growing energy and computational needs.  While some HPC functions cannot be easily interrupted, and thus require consistent, even backup power at the ready, not all computational jobs require this highest level of Quality of Service (QoS) for time and energy, and many can be paused, delayed, or even discarded without undermining the overall computational goals. To this end, we exploit massively parallelizable Monte-Carlo (MC) and other AI jobs comprising up to 10\% of HPC load.  These jobs will provide sufficient initial “Spinning Demand” throughout the grid, creating breakthrough opportunities for grid planning, stabilization, efficiency, GHG reduction, and renewable adoption.  Inclusion of other types of HPC, and even integration with other computational loads are discussed below.  

\subsubsection{Progression of Compute Loads}
We envision the inclusion of HPC computation jobs will move from simple to more complex, starting with HPCMC jobs as discussed above. These HPCMC jobs, pervasive in epidemiology, weather, scientific computing, and machine learning model optimization and validation, share properties of: a) being parallelizable into independent, simultaneously executed jobs, b) lacking strong QoS requirements (on scale of minutes, hours, or even days), and c) lacking strong privacy or security requirements.  As the Sustainable Grid Marketplace grows, inclusion of more complicated, yet parallelization computational loads are envisioned. Federated Learning, and even distributed large language model learning, are examples of AI computational loads that benefit from parallelization, but have higher QoS and inter-process communication requirements that we will consider as the sustainable grid progresses.


\subsubsection{Integration of Compute Loads: Learn While Mining}
\begin{figure}[htbp]
\centering
\includegraphics[width=\linewidth]{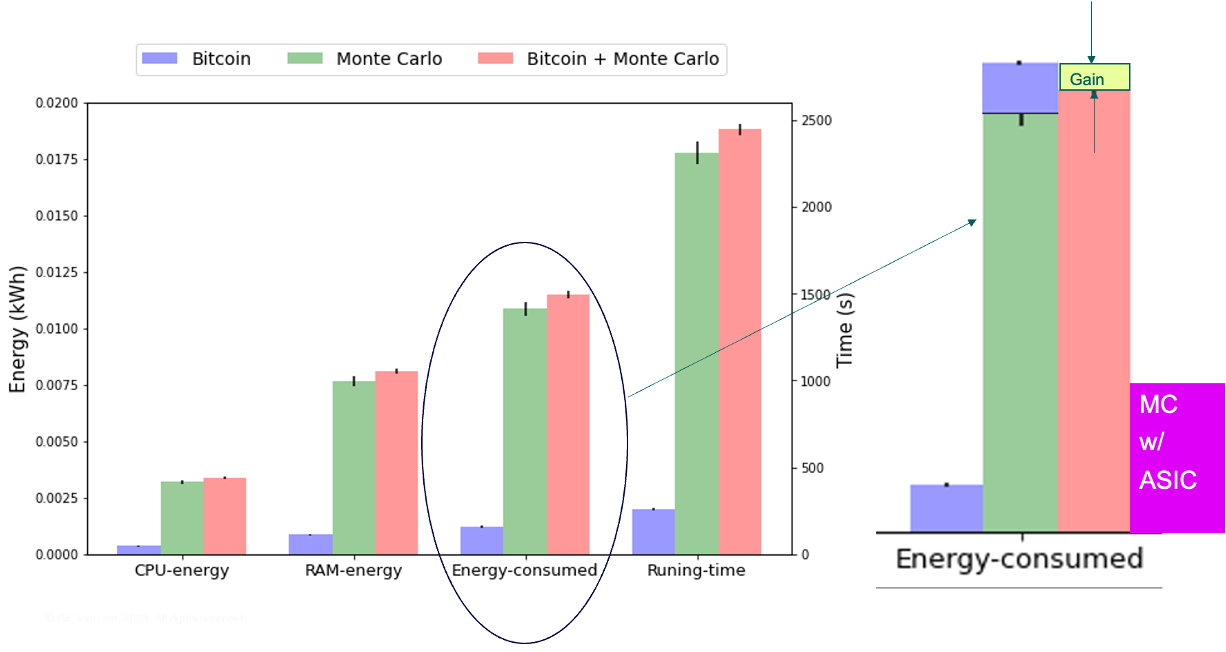}
\caption{Energy benefit of joint Monte Carlo and proof of work mining}
\label{figLearnWhileMining}
\end{figure}

A computational load that is even less complex than HPCMC is proof of work crypto mining. These loads are ubiquitous, and already being used for load leveling activities, but have been resisted due to their high energy cost.  One way the carbon footprint of proof of work crypto mining can be improved is by leveraging it for scientific machine learning as discussed in \cite{evansPatent2024}.  Monte Carlo and Proof of Work computational loads share in common the need for massive amounts of pseudo-random numbers in order to function.  One can think of proof of work crypto mining as an extremely aggressive form of rejection sampling.  The inefficiency of proof of work crypto lies in the fact that the only value add done from the terahashes of random numbers that are generated is the rare authentication of a transaction.  In Fig. \ref{figLearnWhileMining} we show the energy benefit of harvesting the random numbers generated from proof of work mining to simultaneous MC simulations.  Here we see a 10\% energy gain from simultaneous ``Learn While Mining" over separate mining and MC simulations.  Thus jointly conducting these and other types of computational jobs can lead to energy benefits.  

\subsection{Coordinated Load Placement}\label{sec:coord_load}

    Rather than “Demand Response,” which reacts to loads and responds to some objective, our approach will “Actively Place,” and serve at its discretion HPCMC load, leveraging the near zero cost of moving data and compute jobs over the internet to reduce or remove infrastructure requirements. This paradigm leverages the fact that moving energy is very expensive, at \$41.5/MWhr \cite{b4} for 1000 miles, whereas moving information is essentially free \cite{b6} depending on the platform. Thus, the most affordable way to optimize the intermodal energy superhighway is to transport demand using the information grid (the internet), and leverage a portion of the growing demand from HPC to optimize and stabilize the grid itself. Specific metrics measuring progress in economic and security priorities for the US including the provision of energy technologies that reduce GHG emissions, improving efficiency of virtually all economic sectors as the AI revolution continues, and increasing the resilience and reliability of energy infrastructure are as described in Table~\ref{tab1}.
\begin{table}[htbp]
\caption{Expected Benefits}
\begin{center}
\begin{tabular}{ |p{2.5cm}|p{1.9cm}|p{2.8cm}|  }
\hline
\textbf{\textit{Metric}}& \textbf{\textit{State-of-the-art}}& \textbf{\textit{Target}} \\
\hline
 Stability   & Freq Regulation    & 30\%+ Cost Reduction\\
 \hline
Grid Expansion   &+1x    & +0.5x \\
 \hline
Renewables curtailed   & 2-6\%   & $<0.5$\% (~10x reduction)\\
 \hline
HPC GHG Emissions &.86lb/kWh   & .75lb/kWh\\
 \hline
\end{tabular}
\end{center}
\label{tab1}
\end{table}
\subsection{HPC and Grid Stability Support}\label{sec:GridStab}
The most fundamental operation in maintaining grid reliability is balancing supply and demand. Indeed, at any given time the power demand from the load is required to be satisfied by the generation otherwise the grid frequency deviates from its synchronous value (e.g., 60Hz in North America, 50Hz in Europe) which can ultimately lead to grid collapse if not corrected rapidly. In the era of high penetration of HPC and variable renewable energy (VRE) resources such as wind and solar, the grid stability can be significantly challenged by the high volatility of both load and generation. The high dynamics of HPC can be turned into an opportunity to support the grid by providing ancillary services that will increase grid stability while reducing the size and cost of ancillary equipment otherwise required. The grid regulates frequency using two key mechanisms: regulating reserve, including autonomous dispatch or the so-called automatic generation control (AGC), and contingency reserve, including both spinning and non-spinning reserve \cite{EPRI2019}. Regulation and contingency reserves differ essentially by their time scale requirements and the opportunity and magnitude of their intervention. In the US, regulation reserve (also known as primary frequency control) is a continuous operation (non-event) that deploys a dispatch command to generators every 4-6 secs, while contingency reserve is an event-based operation. Spinning reserve (already connected resources) and non-spinning reserve (ready to connect upon request) are required to supply power usually within 10 minutes to help the frequency return to nominal value upon a significant event (e.g. generation trip). Demand response is another form of contingency reserve. Another key difference between regulation and contingency reserve is the directionality of the power contribution. While regulation reserve can increase or decrease the power based on the received control signal (up and down regulation, respectively), contingency reserve is expected to only inject power to restore a collapsing frequency. This sets the boundary of where HPC can contribute to support the grid.
HPC provides the ability to stack compute load and build an instantaneous power demand to help stabilize grid frequency. If this computational power can be modified in real-time to go up and down and follow an automatic dispatch signal such as AGC, they can become a very powerful asset to provide frequency regulation service to the grid. However, regulation reserve service has strict rules to be complied to for eligibility. The most critical include the capability range for ramp up and ramp down which may or may not be symmetrical, ability to follow AGC setpoints adjusted every four to six seconds, minimum duration capability (i.e., minimum duration time for same response direction) which is typically 15 minutes. With future hyperscale data center size ranging from tens of MWs to gigawatts \cite{IBM2024}, HPC have the advantage of size (up to several hundreds MW), modularity (adjustable up and down range from same HPC), and high flexibility. Table \ref{tab2} summarizes key differences between the regulation reserve and secondary contingency reserve and the eligibility of HPC for each.
\begin{table*}[htbp]
\caption{Key grid reliability services and eligibility of HPC}
\begin{center}
\begin{tabular}{ |p{4cm}|p{4cm}|p{4cm}|p{4cm}|}
\hline
\textbf{\textit{Service}}& \textbf{\textit{Regulating Reserve}}& \textbf{\textit{Primary Contingency Reserve}}& \textbf{\textit{Secondary Contingency Reserve}} \\
\hline
 \textbf{\textit{Main function}}   & Continuous Frequency regulation   & Demand Response & Spinning and Non-spinning reserve\\
 \hline
\textbf{\textit{Trigger}}   &Autonomous dispatch (AGC) & Real-time dispatch or Major Event & Major event (e.g loss of generation)\\
 \hline
\textbf{\textit{Response type}}   & Upward and Downward & Downward & Upward\\
 \hline
\textbf{\textit{Response time}}  &Follow setpoint every 4 to 6 secs   & Within 10min & Within 10min\\
 \hline
 \textbf{\textit{Capability basis}}  &5min Ramp Up/Down & Undefined & 10-min Ramp\\
  \hline
 \textbf{\textit{Minimum capacity duration}}  &15min (typical) & Minutes to several hours & 30-60min\\
   \hline
 \textbf{\textit{HPC qualification}}  &Very Likely & Very Likely & Possible\\
   \hline
\end{tabular}
\label{tab2}
\end{center}
\end{table*}

\begin{figure}[htbp]
\centering
\includegraphics[width=\linewidth]{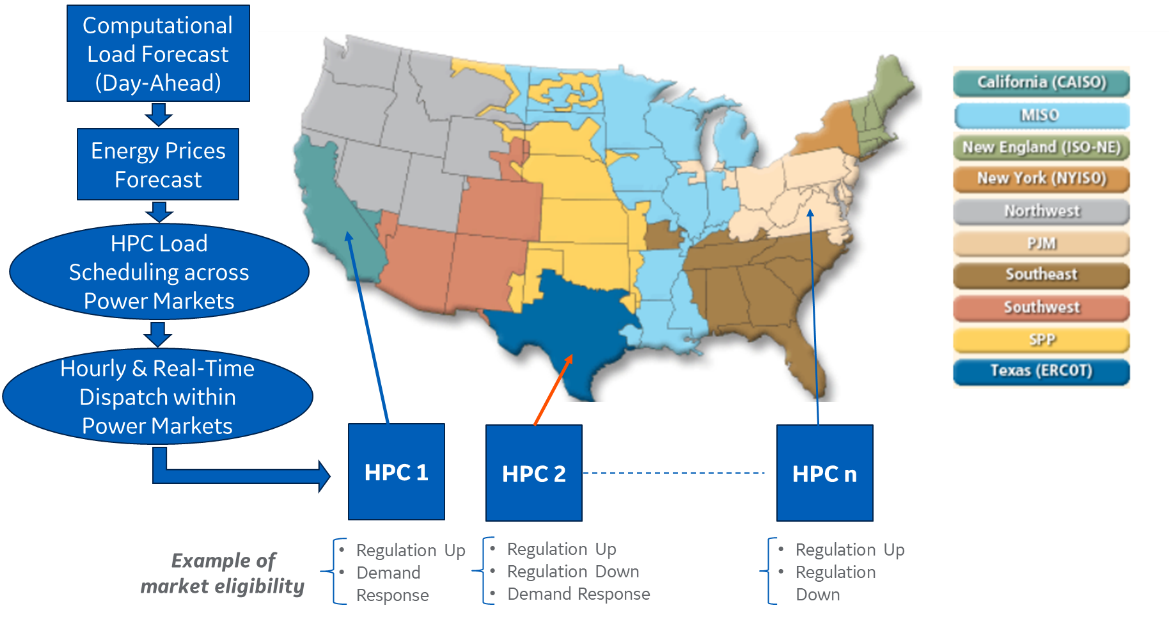}
\caption{Potential utilization of HPC in US electric power markets}
\label{figGridStab}
\end{figure}

The ability of HPC as an asset for regulation reserve is clear. However, its eligibility to participate in contingency reserve such as the secondary frequency regulation is less straightforward in current electricity market practices. Indeed, the contingency reserve requires eligible resources to take rapid action and inject power to the grid upon request to restore the grid frequency to nominal value. Unlike battery energy storage, HPC is a purely “load” resource i.e. it can only absorb power from the grid; therefore, power injection is not an option. However, if the consequence of injecting power under contingency is to reduce the imbalance within an area control and avoid underfrequency load shedding (UFLS), it is conceivable that a rapid ramp down of large flexible load such as HPC can presumably provide the same result. If not readily eligible for “spinning” reserve, HPC can at least be qualified for demand response. Demand response is often considered as a form of secondary contingency but due to eligibility requirements such as its autonomous characteristic, required response time, or directionality, it is effectively a primary frequency response mechanism that can suit very well the capabilities of HPC. Although the participation of HPC in demand response will not impact the system inertia as with traditional spinning and non-spinning resources (e.g. gas or hydro plants), its potential fast response can improve the ROCOF (rate-of-change-of-frequency) and ultimately the frequency nadir.      
The biggest challenge for HPC in providing ancillary services and offsetting grid investment to lower the cost of regulation and contingency reserves is understanding the market rules. Indeed, in many parts of the world, the grid is segmented into regions or balancing authorities administrating the energy markets and the frequency regulation mechanisms. In the US for example, those will be the ISOs (Independent System Operators). Each ISO, due to their generation mix, grid inertia, marginal energy price, etc. will have specific rules affecting the eligibility of HPC to provide ancillary services for grid stability. If forecasting computational load across a region (e.g across US or Europe) allows HPC participants to share the load, the load stacking towards building a profile modifiable in real-time to respond to an AGC signal or a contingency event will be done at the transmission system operator (TSO) level. Fig. \ref{figGridStab} shows a notional segmentation of distributed HPC load that can support participation in frequency markets in different TSO. Each TSO as highlighted in Fig. \ref{figGridStab} will have different market rules and eligibility \cite{FERC2024}. Eligibility requirements and rewards will not be the same per region. In Fig. \ref{figGridStab}, HPC $n$ is an aggregation of regional HPC participants. An additional HPC optimal dispatch within the ISO, constrained or incentivized by ramping capability requirements and other economic and security criteria will need to be developed to appropriately schedule the computational demand for real-time operation. 

Despite the complexity of grid ancillary service mechanisms, the trend for modernizing grids is a larger utilization of ancillary services due to greater volatility of both the generation and load. 
In this context, distributed HPC load is positioned as an important resource, as it is potentially more powerful, cost-effective, and carbon neutral than battery energy storage, aggregated electric vehicles, or hydrogen electrolyzer plants. 

\begin{figure*}[htbp]
\centering
\includegraphics[width=0.85\linewidth]{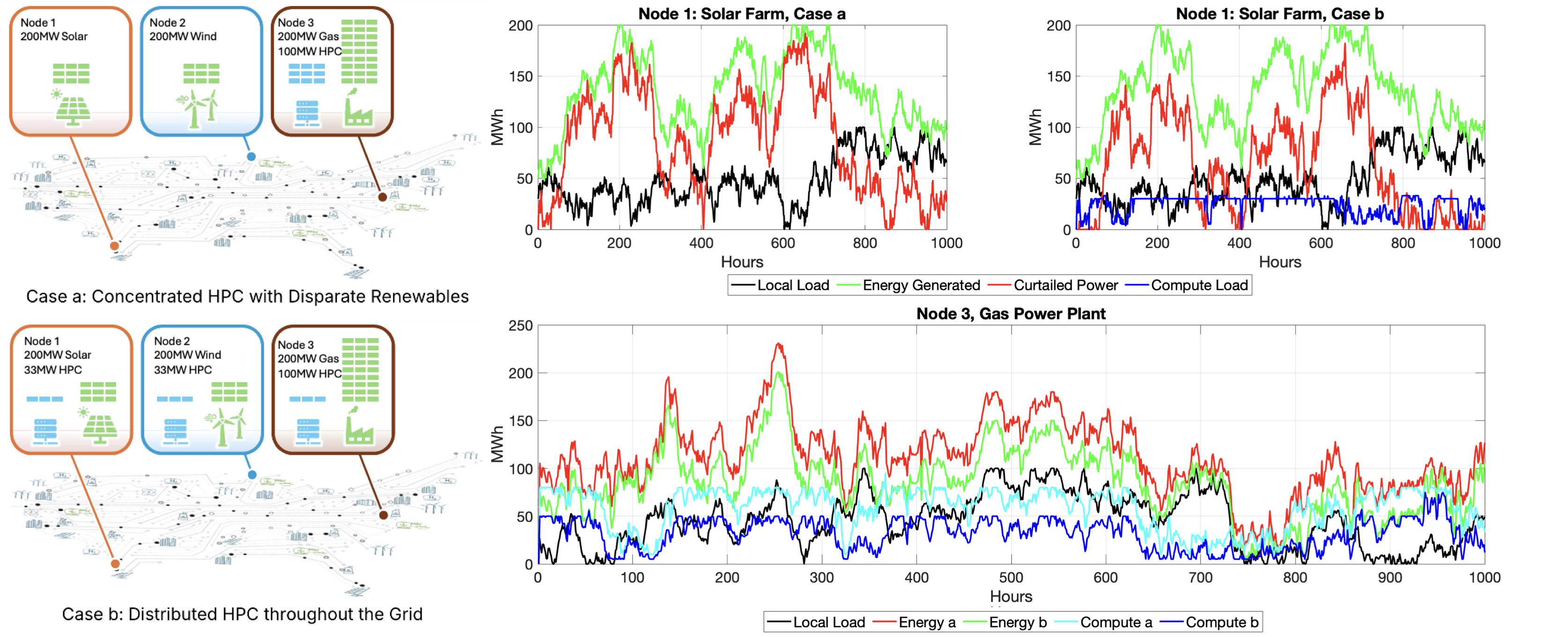}
\caption{(Left) Example simulation HPC configurations. (Right) Energy generation, curtailment, and usage time series in simulation cases a and b.}
\label{fig1}
\end{figure*}
\section{HPC Load Placement \& Scheduling Algorithm and Simulation Results}
\label{sec: Load Optimization Simulation}
We demonstrate the intuition of these concepts via simulation and analysis of a 3-node system. As pictured in the upper portion of Fig. \ref{fig1}, the system includes a solar farm, a wind farm, and a gas turbine plant located at node 1, node 2, and node 3, respectively. We consider two HPC configuration cases: (a) 100 MW HPC concentrated at node 3 alongside the gas turbine (b) 100 MW HPC at node 3 augmented with 33 MW HPC at both node 1 and node 2. Local electricity demand at each node is modeled as a Gaussian random walk with a standard deviation $\sigma=5$, as are wind and solar generation. Compute load arrives at a centralized demand server, to be served across the case-dependent available HPC capacity. Note that we impose generation capacity limits across all nodes as well. Costs for generation and movement of energy, including carbon credit offsets, are described in Table \ref{tabCosts}.
\begin{table}[htbp]
\caption{Cost Assumptions in simulation}
\begin{center}
\begin{tabular}{|p{3.7cm}|p{3cm}|}
\hline
\textbf{\textit{Source}}& \textbf{\textit{Cost (\$/MWhr)}} \\
\hline
Solar Farm (node 1)   & \$10/MWhr\\
 \hline
Wind Farm (node 2)   & \$20/MWhr\\
 \hline
Gas Power Plant (node 3)   & \$50/MWhr\\
 \hline
Energy Transport  $>$ 1000 miles  & \$40/MWhr\\
 \hline
\end{tabular}
\end{center}
\label{tabCosts}
\end{table}

Taking on the perspective of a centralized scheduler, one approach to matching supply to demand is summarized in Algorithm 1. At all nodes, local non-compute demand is met to the extent possible by local generation, with the gas turbine plant acting as a slack bus in scenarios involving renewable shortfalls. Any excess renewable generation is used to serve the centralized compute load, with the gas generator again acting as a slack bus to the extent possible. 
\begin{algorithm}
    \caption{Load Optimization Algorithm}
    \label{algorithm}
    \begin{algorithmic} [1]
    \STATE {\bfseries Input:} Set of Nodes $N=\{n_i, i=1:M\}$ with properties: $E_{Capacity}$, $C_{Capacity}$, and $Carbon_{Intensity}$\\
        \FOR {each time step }
        \STATE $\forall{n} \in{N}$: Calculate excess energy and excess compute capacity for each node and sort by cost
        \WHILE{There exist excess energy at low cost nodes} 
                \STATE Calculate Cost to serve non-compute loads
                \STATE Move Energy as to meet Non Compute Demand, minimizing cost
                \STATE Update excess energy and compute capacity for each node and sort by cost
                \STATE Calculate maximum compute load that can be moved \\
                \STATE Move compute load to lowest cost nodes until compute load met \\
        \ENDWHILE
        \ENDFOR
        \STATE \textbf{Output:} Compute  load transfer profile\\
    \end{algorithmic}
\end{algorithm}

As shown in Fig. \ref{fig1}, curtailed generation levels at the solar farm are higher in case (a) than in case (b) due to the ability to flexibly use excess energy to process compute load, e.g., HCPMC. The lower graph shows the overall impact of distributed HPC on the gas power plant. In case (a) the plant must make up for power deficits in the system, while in case (b) it consistently operates at lower power level, thanks to the distributed compute load serviced by renewable sources. 
\begin{figure}[htbp]
\centering
\includegraphics[width=0.9\linewidth]{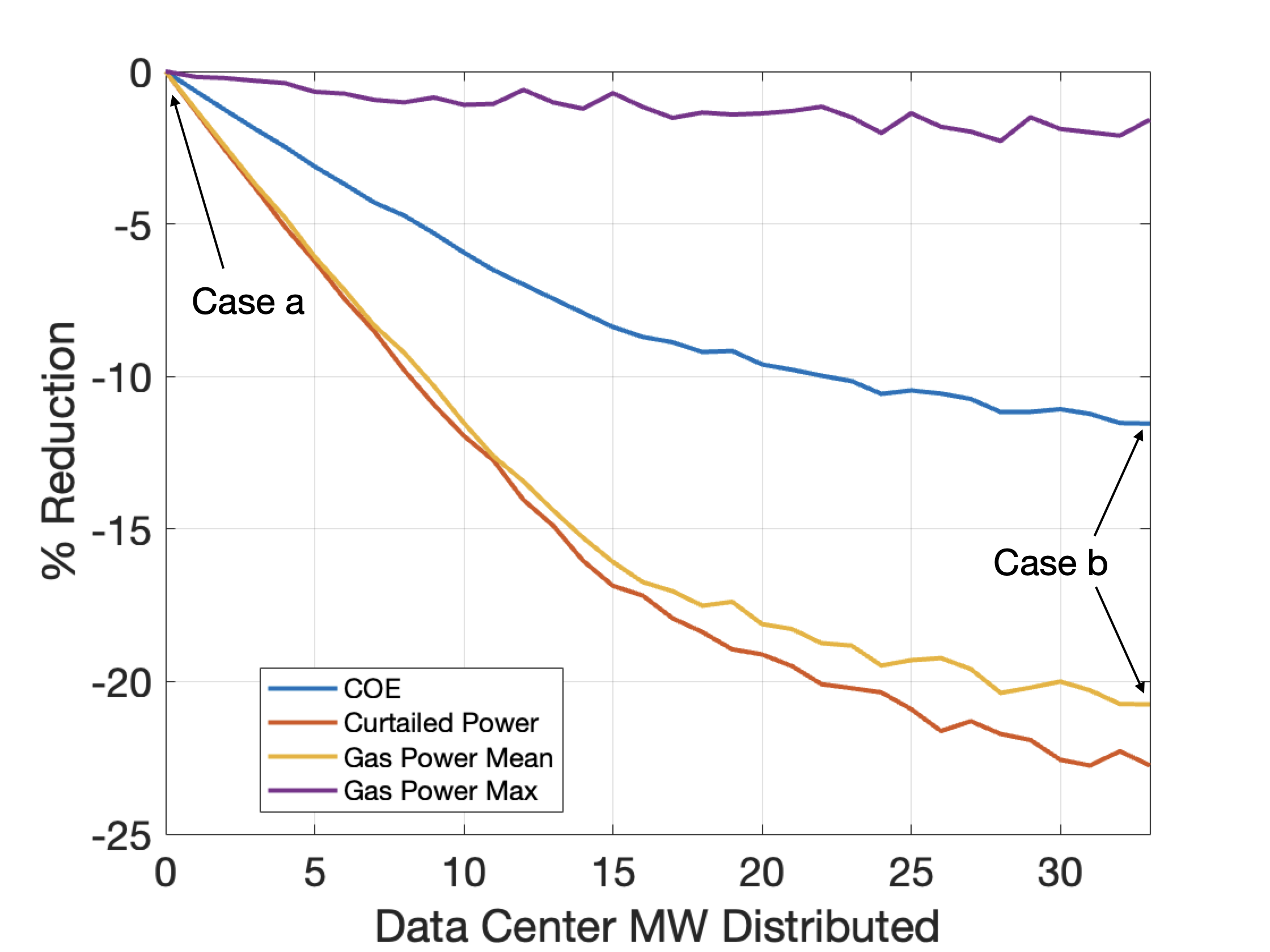}
\caption{Cost of energy (COE), renewable curtailment, and gas power utilization statistics with increasingly distributed HPC.}
\label{fig2}
\end{figure}
An expanded simulation shows the results for varying levels of HPC distribution, shown in Fig.~\ref{fig2} where the simple experiment described above is repeated for 100 bootstraps of Gaussian random walks of power generation and load for levels of distributed HPC between zero (case (a) above) and distribution of additional HPC capacity at the other nodes (33 MW of compute each, case (b)). The cost of electricity, including generation and movement, drops by approximately 12\% as HPC capacity becomes more distributed. Similarly, both curtailed renewable production and mean gas power generation levels decline with more uniform HPC distribution. An additional benefit is the peak gas generator power also drops by 2-3\% as the distribution of compute load increases, implying that 
grid spinning reserve and capacity levels can be reduced.

\begin{figure}[htbp]
\centerline{\includegraphics[scale=0.25]{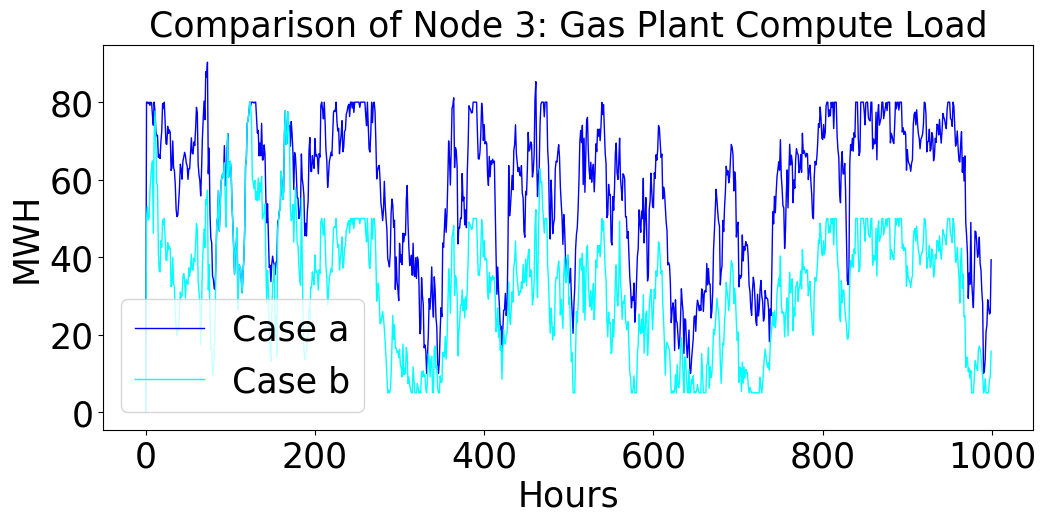}}
\caption{Comparison of gas power plant node compute load for case a and  b.}
\label{fig:comparison}
\end{figure}

Fig. \ref{fig:comparison} further explores nodal operations when the compute load is distributed in a scenario marked by wind farm underproduction. here we observe that with the distributed HPC load among renewable energy sources, compute load served by gas production at node 3 decreases in case (b). This is because we take advantage of curtailed power generated by renewable sources by transferring compute load appropriately.

\section{Market Operation and Pricing}
\label{sec:Market_operation_pricing}

As mentioned in Section \ref{sec:parallel_HPCMC}, a significant portion of compute load is flexible in time. Thus, while HPC capacity may be available at a given moment, some consumers may be amenable to adjusting their usage of these resources depending upon their willingness to pay a given price. On the other hand, renewable producers, particularly those with excess generation slated to be curtailed, may be willing to serve compute load to a degree depending upon prevailing market prices. In this section, we examine a stylized spot HPC capacity market, assessing the impact of HPC distribution across the system described in Section \ref{sec:coord_load}. We assume for simplicity that all HPC capacity in the system is sold in this spot market, after electrical loads are served. Therefore, any renewable adjacent HPC is powered by production that would otherwise be curtailed.

We focus here on the three-node setting of Section \ref{sec: Load Optimization Simulation}
and a market in which the three suppliers each offer supply schedules at each time slot mapping prices to amounts of HPC capacity provided to a single customer, e.g., a compute load aggregator. We assume that compute load arriving at a given time may be only partially served, incurring per unit monetary shortfall penalty $\theta$, and that the consumer solves the following price selection problem at each time step, which balances the cost of compute with a penalty for supply shortfall
\begin{equation}\label{eq:opt_obj}
    \min_{p\geq 0}\quad p\sum_{i=1}^3s_i(p) + \theta(d-s_i(p))_+,
\end{equation}
where $x_+ = \max\{x,0\}$, $p$ is the price selected. We suppress the dependency of compute demand $d$ and the supply functions $s_i$ on time. Regarding unserved compute load at a given time step, we later compare two options: unserved load is either discarded entirely or added to a queue, so that it rolls over and contributes to the amount $d$ in \eqref{eq:opt_obj} at the following time step.

\begin{figure}[htbp]
\centering
\includegraphics[width=\linewidth]{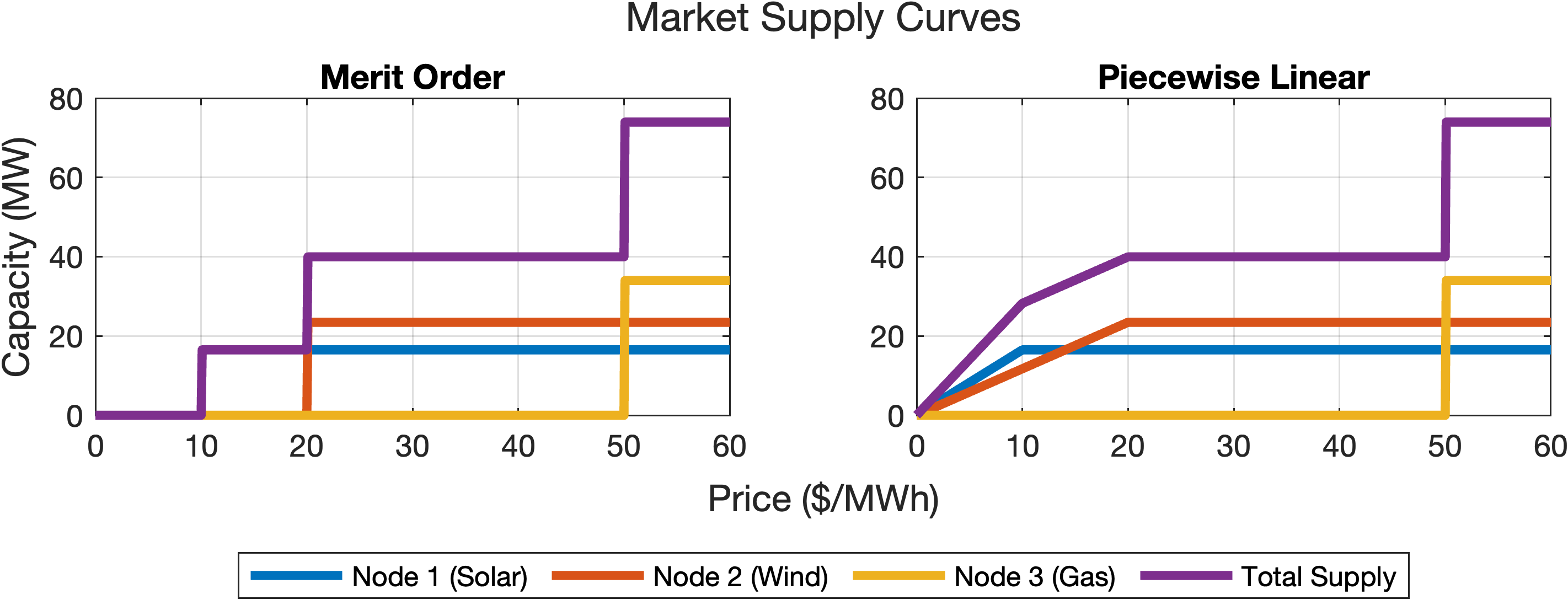}
\caption{Alternative market-wide supply curves arising from merit order or linear supply function bidding from of renewable-backed HPC providers.}
\label{fig:market_curves}
\end{figure}

Turning to pricing, adhering to the well-known merit order supply paradigm for electricity markets \cite{sensfuss2008merit}, one way to price compute jobs as they are distributed geographically is as follows: make use of computing capacity in order of increasing unit cost bids, and set the per unit price equal to the rate charged by the final or marginal supplier. In this setting, bids take the form $s_i(p)=0$ for $p<c_i$, where $c_i$ is a constant marginal cost, and $s_i\in [0,\overline{s}_i]$ for $p\geq c_i$, where $\overline{s}_i$ is the maximum capacity supplier $i$ can provide. The lefthand plot in Figure
\ref{fig:market_curves} shows representative individual and market-wide supply curves under this approach. Note that the flat portions of the renewable supply curves may correspond to a value below the maximum HPC capacity, whenever curtailable energy levels fall below these maximum values. In the context of such bids, $\theta$ essentially sets the customer's highest acceptable price in the sense that no supply will be requested from supply $i$ with $c_i>\theta$.

As we assume that renewable generation is curtailed if not committed to HPC, solar and wind farm operators may be willing to offer HPC capacity for rates below those reported in Table \ref{tabCosts}. One relatively simple alternative bid format to the merit order approach capturing this willingness consists of piecewise linear supply functions (PLSF) specified as linear ramps from zero to the full market rate at each renewable supplier's (time dependent) maximum capacity. The righthand plot in Fig.~\ref{fig:market_curves} shows adjusted supply curves following this format. Note that we assume the gas turbine supply curve is unchanged as production is fully controllable.
Similar curves on the demand side in related settings have been shown to be helpful in improving productivity and lowering costs \cite{islam2017spot}.


\begin{figure}[htbp]
\centering
\includegraphics[width=\linewidth]{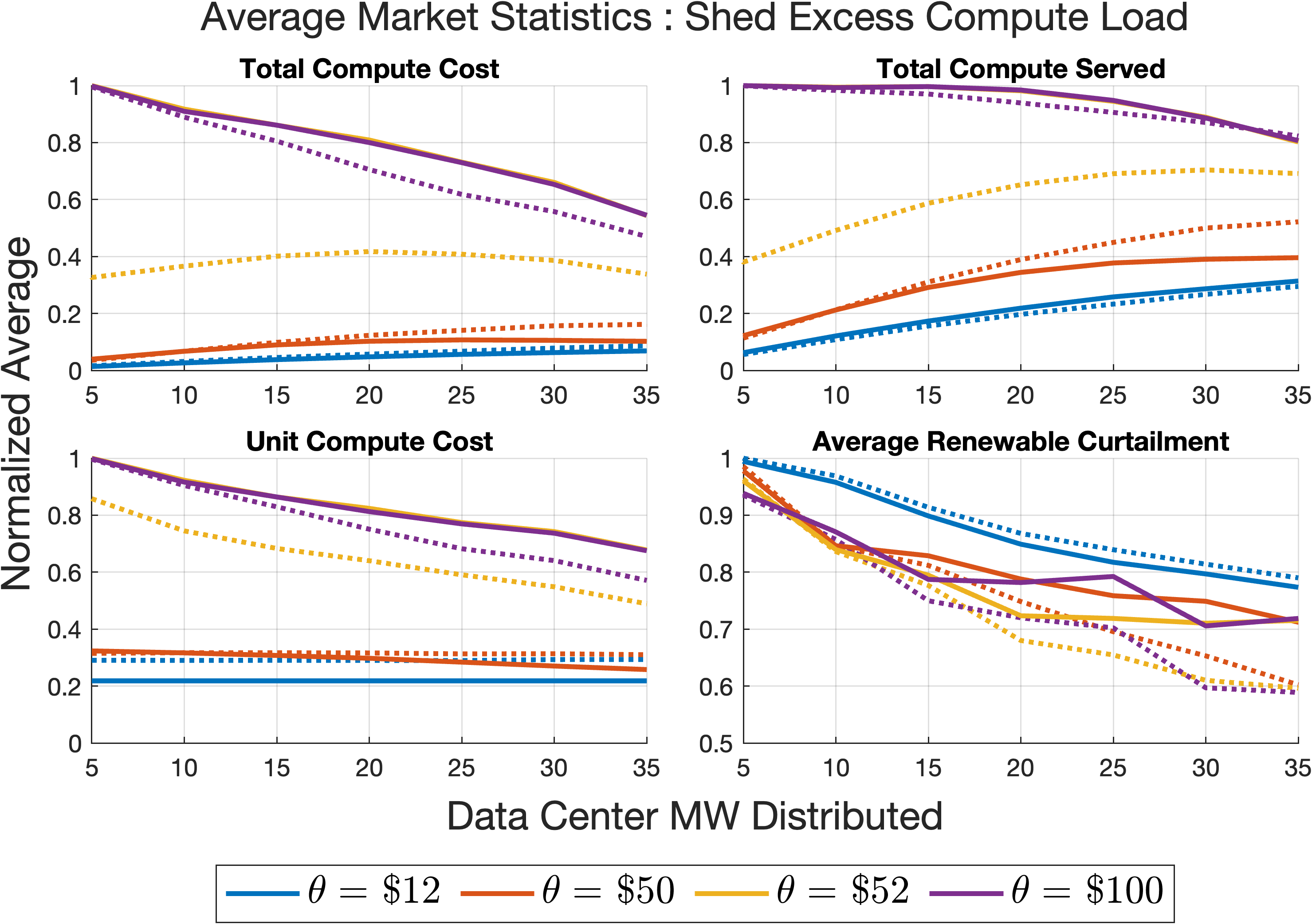}
\caption{Mean metrics over 1000 traces (solid : merit order, dotted : piecewise linear supply functions).}
\label{fig:merit_supply}
\end{figure}
\begin{figure}[htbp]
\vspace{3mm}
\centering
\includegraphics[width=\linewidth]{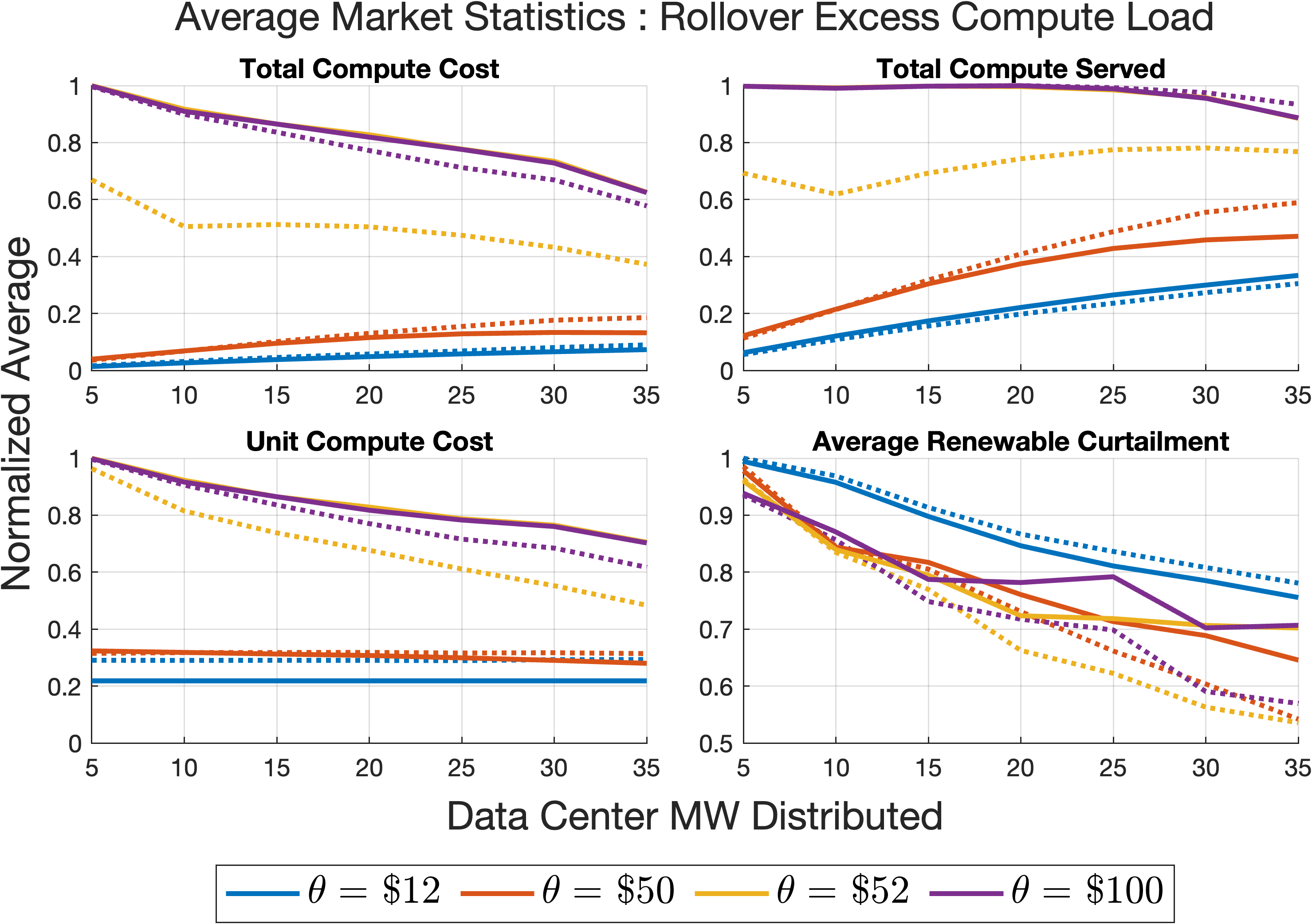}
\caption{Mean metrics over 1000 traces (solid : merit order, dotted : piecewise linear supply functions).}
\label{fig:supply_fcn_bidding}
\end{figure}

Figs.~\ref{fig:merit_supply} and \ref{fig:supply_fcn_bidding} illustrate outcomes associated with these bid formats assuming unmet compute load is shed or rolled forward in time, respectively, for select values of $\theta$, in terms of average compute total compute cost, service and unit cost (overall service cost divided by load served), along with the average amount of renewable production curtailed after commitments are made to HPC. Averages are taken over 1000 traces each, with each trace consisting of 5 minute interval time steps over 24 hours. 
In these simulations, we held the total network HPC level constant at 100 MW, and horizontal axis values in Figs.~\ref{fig:merit_supply} and \ref{fig:supply_fcn_bidding} refer to the capacity located at each node 1 and 2. For instance, at the `30' tick marks, 60 MW of HPC capacity is evenly split between nodes 1 and 2, with 40 MW left to node 3. Each subplot in either figure is individually normalized by the maximum of the raw data underlying the curves displayed. Solid lines correspond to merit order supply, while dotted lines are used for PLSF bids. Fig.~\ref{fig:merit_supply} plots results assuming that any excess compute load remaining after the consumer selects their price point via \ref{eq:opt_obj} is discarded, while Fig. \ref{fig:supply_fcn_bidding} displays results when the consumer simply rolls remaining compute load over to the next slot, until the end of a given 24 hour trace. 

With respect to the parameter $\theta$, in view of Table \ref{tabCosts}, under the merit order approach, the \$10/MWh rate offered by the solar farm represents the lowest $\theta$ value for which \emph{any} HPC capacity can be purchased, while the \$50/MWh rate offered by the gas turbine represents the minimum $\theta$ value required to make use of the entire network's HPC capacity, regardless of location. Note that under merit order market operation, mean cost and service totals are essentially the same even as $\theta$ nearly doubles from \$52 to \$100, while under the PLSF format, a tradeoff is evident: lower willingness to pay translates to lower total compute served, at lower total and per unit costs. While not plotted, a continuum of such curves exists for the PLSF setting, both between the \$12 and \$50 curves, as well as the \$52 and \$100 curves, reflecting greater flexibility in terms of how the market responds to varying customer preferences. 

At a high level, both Figs.~\ref{fig:merit_supply} and \ref{fig:supply_fcn_bidding} demonstrate that PLSF bids yield reduced average renewable curtailment, particularly at high levels of renewable HPC colocation. Assuming a high willingness to pay (or equivalently, eagerness to meet demand quickly) on the part of the HPC customer, PLSF bids offer comparable total compute service levels at reduced total and per unit costs. For lower values of $\theta$, overall and per unit costs increase slightly under PLSF bids, while compute load served increases as $\theta$ approaches \$50. 

Concentrating on the comparison between shedding or rolling over excess compute load at the end of a given time slot, the latter yields higher levels of average total compute served, and lower levels of renewable curtailment, particularly for high values of $\theta$. In terms of compute flexibility in time, these two approaches represent outcomes under two extremes - either compute load is to be served immediately, or it can essentially be delayed indefinitely, or at least to the end of a particular trace (at cost). We leave deadline-aware scheduling scenarios falling between these two extremes for future work.  



\section{Cost Benefit Analysis}
\label{sec:Cost_benefit_analyis}
The California Independent System Operator’s (CAISO) weighted real-time average prices in 2022 were \$22.98 per/MWh for regulation down, \$12.53/MWh for regulation up, \$8.51/MWh for spinning reserve, and \$2.08/MWh for non-spinning reserve \cite{CAISO}. By utilizing renewable assets at max capacity, and providing schedulable real-time variable loading, utilities can release a controllable percentage of their generational capacity to the grid downstream of the schedulable load.

This paradigm allows these assets to operate on very short time scales as spinning reserve as well as regulation up or down within operational bounds that do not involve the dynamics or time scales of the turbines.  Adding this utility-controlled demand response allows for operating assets at close to steady state and shaping the output to the grid based on the relative energy cost of selling output, or consuming it to sell as compute capability.  Co-located HPC allows for the introduction of a salvage value for otherwise curtailed power, capitalizing on the opportunity cost of previously unutilized generation capacity to provide compute as a service.  CAISO curtailed 2.4 TWh in 2022 that otherwise could have been generated and re-directed to revenue generating compute services rather than being curtailed \cite{Solar2023}.

Within the current grid control paradigm, marginal increases in renewable power installations create commensurate increases in the curtailment rate.  By introducing controllable loading, this paradigm can be flipped, reducing the curtailment rate with the introduction of net demand that is coordinated by the utility and optimized based on marginal energy costs.  Based on a study from UC Davis, an estimated marginal 1 MWh increase in hourly net demand added in California results in a reduction in the average daily curtailment rate for utility-scale solar and wind, with a marginal reduction in the curtailment rate of 1.5\% for this newest MWh of load.  With additional granularity for time of day, scheduling this marginal demand for times of peak effect (10am in this case), one could expect as much as a 5.3\% reduction in marginal curtailment rate \cite{Novan2023}.  The effect of this co-located compute would be beneficial to the gas industry as well.  Allowing turbines to operate above capacity with downstream load to filter its output.  This method can have a smoothing effect on the ramp rates necessary to meet variable demand helping to extend the lifetime of a turbine.

\section{Conclusions and Next Steps}
\label{sec:conclusions_and_next_steps}

We plan to extend this work to the deployment of HPC loads towards grid stability related services including voltage and frequency regulation using test systems, e.g., IEEE benchmarks, and expand generation modeling to include coal and other carbon intensive sources. Accurate grid modeling and simulation will be essential to analyze the grid's behavior under various energy portfolios and load conditions (e.g., local and compute load) when coupled with our HPC scheduling approach. 
Regarding market design, we plan to incorporate demand-side bidding and examine two-sided market clearing, including equilibrium analysis and strategic behavior, and consider issues around compensating flexibility, as well as the role of HPC workload aggregators. 
In total, we aim to build a comprehensive approach to efficiently maintaining grid stability and reliable power delivery that leverages spinning HPC demand. 

\bibliographystyle{IEEEtran}
\bibliography{reference}

\end{document}